# The Quiet and the Compliant: How Regulation and Polarization Shape Conventional Wisdoms on Corporate Social Engagement in High-risk Settings

Prof. Jason Miklian
Centre for Global Sustainability
University of Oslo (Norway)
jason.miklian@globe.uio.no

**Abstract**

How do multinational corporations govern social engagement in fragile and conflict-affected settings, and what do the perceptions of the professionals who design these initiatives reveal about the institutional forces shaping corporate social purpose in crisis contexts? We present a synthetic survey of 400 corporate professionals in Europe and the United States working on social impact in high-risk environments, introducing a "conventional wisdom baseline" methodology that formalises what existing theory predicts as a benchmark for future field testing. Drawing on international business, political CSR, and business-peace literatures, we test seven hypotheses about how regulatory environments, political polarization, sector characteristics, and headquarters-subsidiary dynamics shape corporate social engagement. European professionals report significantly higher strategic integration of social impact across all measured dimensions, consistent with an institutional distance interpretation in which mandatory due diligence restructures MNC organizational logics abroad. US professionals overwhelmingly report that political polarization hinders social initiatives, yet this perception does not predict unreported social activities, complicating the "quiet CSR" narrative. Extractive industry professionals display both the highest operational preparedness and the highest complicity awareness, a pattern we term presence-dependent reflexivity. We introduce the concept of constitutive tension to describe how mandatory due diligence reconstitutes rather than supplements organizational logics, and identify a measurement void where ESG frameworks fail to capture corporate performance in fragile contexts. These patterns establish theoretical propositions for empirical testing.

**Keywords**: corporate social responsibility in fragile states; headquarters-subsidiary governance; mandatory due diligence; ESG measurement validity; synthetic survey methodology

## Introduction

The growth of corporate social engagement in fragile and conflict-affected settings is one of the most consequential yet least understood developments in the governance of global business. Over the past two decades, a substantial body of scholarship has examined whether private sector actors can contribute to peace and social welfare in places where state capacity is weak, violence is endemic, and institutional frameworks are contested (Fort, 2007; Ganson & Wennmann, 2016; Miklian & Schouten, 2018). A parallel literature on political CSR has theorised the conditions under which corporations assume quasi-governmental roles in filling institutional voids, with particular attention to the normative and democratic implications of that assumption (Scherer & Palazzo, 2011; Scherer et al., 2016). Yet these conversations have proceeded in relative isolation from a third body of work on ESG measurement, ratings divergence, and the political contestation of sustainability (Berg et al., 2022; Curtis, 2025; Tang et al., 2024).

The result is a fragmented theoretical landscape in which the actors who design and implement corporate social initiatives in difficult operating environments remain empirically invisible. This gap speaks directly to international business research. How MNCs govern social engagement when operating across institutional contexts marked by weak state capacity, violence, and regulatory contestation is a question about institutional distance (Xu and Shenkar 2002), headquarters-subsidiary governance (Kostova and Roth 2002), and how home-country regulatory mandates reshape organizational logics abroad. The IB literature has examined subsidiary governance, knowledge transfer, and institutional adaptation extensively, but rarely in settings where the host-country institutional environment is not merely different from the home country but actively dangerous.

The professionals who manage corporate social impact in fragile contexts occupy an organisational position that is structurally challenging to scholarly observation. They work at the intersection of compliance, strategy, operations, and public affairs, often across multiple jurisdictions, and their decisions are shaped by regulatory mandates, political pressures, security conditions, and organisational hierarchies that vary across firms, sectors, and headquarters locations. Existing empirical approaches, whether based on disclosed reports, ESG ratings, case studies, or executive interviews, capture fragments of this professional reality without reconstructing the perceptual world in which strategic choices are made. The few comparable survey efforts that exist focus on sustainability professionals in stable contexts (BSR/GlobeScan, 2019) or on risk managers responding to geopolitical threats (International SOS/Ipsos MORI, 2023), but none have targeted the specific population of corporate actors working on social impact in settings characterised by crisis, conflict, and operational fragility.

I address the importance of filling this gap by conducting a synthetic survey of 400 professionals working at headquarters locations for firms with over 1,000 employees on corporate social impact in fragile and crisis-affected settings, split between the United States and Europe. The survey captures eight demographic and professional screeners alongside 23 Likert-scale items measuring perceptions of strategic integration, regulatory pressure, political polarization, operational preparedness, supply chain coordination, subsidiary autonomy, and ESG ratings validity. Our theoretical framework is grounded in international business research on MNC governance across institutional contexts, drawing additionally on political CSR and business-peace scholarship. From these, I derive seven hypotheses that predict how regulatory divergence between the EU's mandatory due diligence regime and the US's contested approach to ESG should manifest in measurable differences across professional perceptions and practices.

There is a growing regulatory divergence at the centre of this framework. The European Union has moved toward mandatory sustainability due diligence through a series of overlapping instruments culminating in the Corporate Sustainability Due Diligence Directive (CSDDD). The United States has moved in the opposite direction: by early 2026, more than 480 anti-ESG bills had been introduced across state legislatures, and several major institutional investors had withdrawn from climate-focused coalitions under political and legal pressure. These trajectories create what approximates a natural experiment (complicated by pre-existing institutional differences, as discussed in limitations) in the relationship between regulatory environments and corporate social purpose in high-risk settings. I exploit this divergence to test whether mandatory due diligence reshapes organisational logics beyond compliance, whether political polarization produces strategic silence alongside reduced engagement, whether sector-specific modes of presence in fragile contexts generate distinctive patterns of operational preparedness and self-critique, and whether scepticism about ESG measurement frameworks correlates with, or is independent of, operational maturity.

A methodological contribution underpins the empirical design. This paper introduces a 'conventional wisdom baseline' approach to survey-based research: researcher-constructed synthetic data that formalise what existing theory and empirical evidence predict, creating a structured benchmark against which field data can later reveal where professional realities depart from scholarly expectations (redacted). The results reported here are based on synthetic data generated through this approach to validate the survey instrument, test the analytical framework, and calibrate hypothesis testing procedures. The epistemic status of synthetic data requires careful handling: the patterns recovered from constructed data demonstrate that the instrument and analytical strategy can in principle detect theorised dynamics, but they do not constitute empirical evidence. The distinction between designed patterns (those specified in the data generation process) and emergent patterns (those that arose from the interaction of specified parameters) is critical, as emergent patterns carry greater diagnostic value for field deployment.

These findings advance four lines of theoretical development. First, mandatory due diligence may operate not merely as a compliance mechanism but as a force that restructures how organisations understand their social purposes, generating new capacities and new frustrations in the same regulatory movement; I term this constitutive tension, aligning with established concepts of institutional complexity (Greenwood et al., 2011) and institutional work (Lawrence et al., 2009). Second, there may be a growing political economy of corporate silence in which US firms conduct more unreported social activity than EU firms, but this pattern is not driven by perceived political hostility, complicating the "quiet CSR" narrative that has gained currency in practitioner discourse. Third, a null finding on ESG ratings validity may prove consequential: professionals who recognise the limitations of ESG frameworks in fragile contexts are no more operationally sophisticated than those who trust them, suggesting that scepticism and maturity may be independent dimensions of professional experience. Fourth, sector-level variation is presence-dependent reflexivity: the physical and relational depth of a firm's engagement in fragile contexts structures its capacity for self-criticism, as extractive industry professionals carry both the highest operational preparedness and the highest complicity awareness.

## Theoretical Framework and Hypotheses

The scholarly landscape on corporate social engagement in high-risk operational settings presents a striking paradox. Work on business in peace holds that private sector actors can, and perhaps should, contribute to social welfare in places where governance failures leave communities vulnerable (Fort et al. 2024). Critical scholars, in turn, have documented how corporate presence in fragile contexts can intensify conflict dynamics, entrench elite capture, and amount to reputational window dressing (Jamali et al., 2017; Kim & Davis, 2016). Yet neither framing quite captures the complex, often contradictory reality of how firms conceive of, design, and implement social purpose initiatives in places marked by crisis, insecurity, and institutional fragility.

Most empirical work on this examines corporate behaviour from the outside, through disclosed reports, case studies, or ratings, rather than capturing the perceptions and strategic reasoning of the professionals who make these decisions from within. The few comparable efforts, including the BSR/GlobeScan State of Sustainable Business surveys (conducted 2012 to 2019, now discontinued) and the UN Global Compact/Accenture CEO studies, focus on sustainability writ large rather than the challenges of operating in high-risk environments. Siltaloppi et al. (2021) and Rettberg (2006) have examined strategic integration through a tension management lens, but their analyses draw on case studies of individual firms rather than systematic cross-national data. The result is a literature rich in normative prescription but thin on systematic evidence regarding how corporate actors understand the trade-offs and institutional pressures that shape social engagement in fragile contexts.

Therefore, this paper delivers a new synthetic survey of 400 corporate social impact professionals engaging with conflict and crisis settings of operation, split between the United States and Western Europe. Our theoretical framework is rooted in international business scholarship on how MNCs govern social engagement across institutional contexts, integrating insights from political CSR (Scherer & Palazzo, 2011; Scherer et al., 2016), institutional voids theory (Khanna & Palepu, 1997; Jamali & Karam, 2018), and the business-peace literature (Katsos & Miklian, 2024; Joseph et al. 2025). The IB tradition offers analytical resources that neither political CSR nor business-peace provides on its own: a focus on how home-country institutions shape MNC behaviour abroad, on the governance of headquarters-subsidiary relationships, and on how firms manage the tension between global integration and local responsiveness (Bartlett and Ghoshal 1989). The institutional work perspective (Lawrence et al. 2009; 2011) also examines how actors create, maintain, and disrupt institutional arrangements, and with paradox theory in management (Smith & Lewis, 2011), which analyses how organisations navigate contradictory demands simultaneously. I develop seven hypotheses about how regulatory context, sector characteristics, and organizational maturity shape the ways firms approach social purpose in their most challenging operating environments.

*Regulatory divergence and strategic integration*

A consequential structural difference shaping corporate social engagement is regulatory divergence between the European Union and the United States. The EU has moved toward mandatory sustainability due diligence through overlapping instruments: the French Duty of Vigilance Law of 2017, the German Supply Chain Due Diligence Act of 2023, and the Corporate Sustainability Due Diligence Directive (CSDDD), whose Omnibus text was published in February 2026 with an application date of 2029 (Sinnig & Zetzsche, 2024). Given that board diversity and sustainability committee presence enhance human rights due diligence effectiveness (Torelli et al. 2025), these regulatory mandates aim to reshape governance structures and require companies to identify and mitigate adverse impacts across their value chains, introducing civil liability for non-compliance.

The US trajectory is different. While the Dodd-Frank Act's Section 1502 established limited conflict minerals disclosure requirements in 2010, ESG regulation has faced new political opposition. By early 2026, more than 480 anti-ESG bills had been introduced across US state legislatures, with 52 enacted into law, restricting public pension funds from considering ESG factors in investment decisions and contributing to declining ESG investment flows (Tang et al., 2024; Curtis 2025). Non-market strategy literature helps explain how firms respond to such political contestation. Corporate political activity and social engagement are often jointly determined by institutional pressures (Mellahi et al. 2016), suggesting that the anti-ESG movement may reshape the underlying strategic logic of social engagement. EU-headquartered firms are thus compelled to integrate social due diligence into corporate strategy, while US firms face a polarized environment where such integration may attract political backlash.

Therefore, mandatory due diligence regimes may reshape organizational logics, embedding social impact considerations into strategic planning, operational decision-making, and cross-functional coordination in ways that voluntary frameworks do not. Regulatory institutional pressures are among the strongest predictors of substantive CSR adoption (Jamali and Karam 2018), and mandates can transform CSR from a discretionary activity into an embedded governance function, requiring active management of tensions between social purpose and commercial objectives (Siltaloppi et al. 2021). The institutional work perspective (e.g. Lawrence et al., 2009) adds a further dimension: mandatory due diligence does not simply impose external constraints but triggers processes of institutional creation and maintenance within organisations, as professionals develop new routines, metrics, and coordination mechanisms to comply with regulatory demands. This form of institutional work may reshape organisational identities and logics in ways that extend beyond the letter of the law. From an IB perspective, this dynamic reflects how home-country regulatory institutions shape the organizational logics that MNCs carry into foreign operations. The CSDDD does not merely regulate conduct within EU borders; it restructures the terms on which EU-headquartered firms engage with their most challenging host-country environments, creating what Xu and Shenkar (2002) would recognise as an institutionally driven reconfiguration of the headquarters-subsidiary relationship.

Hypothesis 1: *Professionals in EU-headquartered firms will report higher levels of strategic integration of social impact initiatives in high-risk environments than their US counterparts, as measured across items capturing strategy integration (Q2), leadership support (Q6), defined metrics (Q10), and cross-functional engagement between supply chain and sustainability teams (Q19).*

The regulatory environment should also shape how firms perceive the relationship between compliance and authentic social engagement. The CSDDD and its predecessors create a dual dynamic: they compel engagement but also risk bureaucratizing social purpose into a compliance exercise. We would expect EU respondents to report that regulations drive their social engagement and that regulations reduce their flexibility to respond to local conditions. This reflects what paradox theory (Smith & Lewis, 2011) would see as a performing paradox, in which compliance and authenticity coexist as contradictory but interdependent demands rather than resolving into a simple trade-off. The concept of institutional complexity (Greenwood et al., 2011), where organisations face incompatible prescriptions from multiple institutional logics, is also relevant here, though our focus is on the tension within a single regulatory logic rather than between competing ones.

Hypothesis 2: *Regulatory pressure is positively associated with both reported social engagement (Q4, Q12) and perceived constraints on flexibility (Q13), such that firms under mandatory due diligence regimes report higher levels of both engagement and bureaucratic burden than firms operating under voluntary frameworks.*

The U.S. politicization of ESG represents the most significant exogenous shock to corporate social purpose since the concept's institutionalization in 2004. The anti-ESG movement, driven by a coalition of state attorneys general, conservative think tanks, and aligned political actors, has reframed corporate social engagement from a governance best practice into a contested activity (Tang et al., 2024; Curtis, 2025). This contestation has no precedent in the European context: while EU firms face growing regulatory expectations to increase social engagement, US firms face political pressure to decrease or de-emphasize it. US professionals may report higher perceptions of political polarization as an obstacle to social engagement in high-risk environments. EU respondents, though not immune to emerging anti-ESG sentiment, operate within a policy environment where social due diligence retains mainstream institutional support.

Political polarization affects not only whether firms engage but how they communicate. Several US firms have continued or expanded social impact programming while reducing public discussion of these activities, a phenomenon termed "quiet CSR", the "conservative turn", or "greenhushing". Shareholder engagement on ESG issues produces measurable performance improvements, but this analysis generally precedes the anti-ESG movement's peak (Barko et al. 2021). This dynamic should be detectable: US respondents who report that political polarization has made social initiatives harder (Q14) should also report that their firm undertakes social activities not captured in formal reporting (Q8). Q8 may also capture routine organisational slack where social activities fall outside reporting scope, and structural disconnection between operations and reporting functions. The hypothesis tests whether the predicted association exists; its interpretation would require further investigation to distinguish between these mechanisms.

Hypothesis 3: *US-based professionals will report higher agreement that political polarization has made it harder to pursue social initiatives in high-risk environments (Q14) than EU-based professionals. Among US respondents, agreement on Q14 will be positively associated with agreement that the firm undertakes unreported social activities (Q8), reflecting strategic opacity in response to political contestation.*

*Sector effects: operational proximity and the complicity question*

Not all firms experience fragile contexts in the same way, and sector-level variation should produce differences in how professionals perceive their firm's social role. The business and peace literature emphasizes that a firm's mode of engagement with fragile contexts, whether through direct physical presence, supply chain relationships, or financial intermediation, shapes both its potential for positive impact and its risk of complicity in harm (Ganson & Wennmann, 2016). Extractive industries maintain the most direct and consequential physical presence in fragile environments. To wit, 80% of reporting corporations can not determine the country of origin of conflict minerals in their products, and only 1% certify conflict-free status (Kim and Davis 2016). Headquarters-level CSR violations produce negative effects on subsidiary performance, with greater intensity in high-risk operating environments (Nuruzzaman et al. 2024). Extractive firms also face the starkest tensions between operational continuity and escalating conflict: decisions about whether to stay, scale back, or exit when violence intensifies are existential rather than abstract.

I expect these realities to produce a distinctive response pattern among extractive sector professionals, characterized by higher operational maturity on some dimensions (grievance mechanisms, security information flows, exit criteria) alongside greater willingness to acknowledge that their firm's presence can worsen conflict dynamics. Standard business decisions can reconfigure local stakeholder networks in ways that escalate conflict, and extractive sector professionals, by virtue of their proximity to these dynamics, may be more likely to recognize this than peers in sectors with less direct exposure.

Hypothesis 4: *Professionals in extractive industries will report higher levels of operational preparedness for conflict escalation (Q9, Q15, Q16) than professionals in technology, financial services, or consumer goods sectors, but will also report higher agreement that their firm's operational presence can worsen social or conflict dynamics (Q22).*

Supply chain-intensive sectors like consumer goods, retail, and agribusiness engage with fragile contexts through sourcing relationships rather than direct presence. For these firms, the locus of social impact lies in multi-tier supply chains, and a critical question is whether supply chain management and sustainability functions are integrated enough to address these challenges. The CSDDD and German LkSG have made this integration a legal requirement for EU firms, but organizational integration is a structural challenge that regulation alone cannot resolve. Governance interventions can improve shareholder value while increasing stakeholder harm reports (Connelly et al. 2025), suggesting that structural integration and substantive performance do not move in lockstep.

Hypothesis 5: *In supply chain-intensive sectors (consumer goods, retail, agribusiness), the degree of integration between supply chain and sustainability teams (Q19) is positively associated with reported strategic integration of social impact (Q2), social outcome metrics (Q10), and engagement with senior management (Q20), forming a coherent organizational maturity cluster.*

*Information asymmetries and subsidiary autonomy*

A less examined dimension of corporate social engagement concerns the relationship between headquarters and local operations. Firms operating in fragile environments face an information problem: local conditions change fast, local staff possess contextual knowledge that headquarters lacks, and the adequacy of corporate social policies depends on whether they can be adapted to local realities. This gap between headquarters-level CSR commitments and subsidiary-level implementation informs "selective decoupling" in which symbolic adoption substitutes for substantive practice (Jamali et al. 2017). Parent company sustainability reporting practices diffuse to subsidiaries, but cultural distance and local institutional context moderate this relationship (Fourati et al. (2025), suggesting that headquarter-subsidiary alignment on social purpose varies with context.

The international business literature on subsidiary governance offers frameworks for understanding these dynamics. As subsidiary adoption of headquarters practices depends on institutional profiles of both home and host countries (Kostova and Roth 2002; Birkinshaw and Hood 1998), autonomy can be perceived as a general construct rather than a deliberate strategic choice in contexts where centralised control is impractical. Firms that grant local subsidiaries greater autonomy to shape social policies should thus report weaker strategic integration from headquarters, larger information gaps between local operations and senior decision-makers, and more unreported social activities. This may reflect pragmatic adaptation to the realities of fragile contexts, or that "strategic integration" and "local responsiveness" operate as competing organizational logics. A negative correlation could also reflect a selection effect: firms may grant more autonomy to subsidiaries precisely in contexts where centralised strategy is impractical, making the association one of context (Ambos et al., 2010). The integration-responsiveness framework (Bartlett and Ghoshal 1989) provides additional analytical purchase: the CSDDD may be shifting EU firms toward greater integration on social governance precisely in contexts where local responsiveness is most operationally critical, producing a structural tension that voluntary frameworks leave unresolved.

Hypothesis 6: *Reported local subsidiary autonomy (Q18) is negatively associated with strategic integration of social impact into corporate strategy (Q2) and the timeliness of security information reaching headquarters (Q16), but positively associated with unreported social activities (Q8), reflecting a structural tension between centralized strategy and localized adaptation.*

*The ESG ratings validity problem in fragile contexts*

The growth of ESG ratings has created a powerful (and popular) institutional infrastructure for evaluating corporate sustainability performance. Yet the ability of these instruments to capture corporate behaviour in fragile contexts is undemonstrated. Berg et al. (2022) documented that correlations across major ESG rating agencies average only 0.61, with half of the divergence attributable to methodology differences, especially actute in the social component of ESG ratings (Bruno et al. 2026). In fact, that firms in the lowest ESG quartile show the strongest performance improvements following shareholder engagement (Barko et al. 2021), suggesting that ESG scores capture starting positions rather than capacity for change.

If the professionals who work in these settings recognize this limitation, then we should observe scepticism about ESG ratings coexisting with greater organizational maturity in social impact management. Firms that invest most in defined metrics, grievance mechanisms, and systematic social engagement may be those most aware of the disconnect between their internal practices and how those practices are evaluated from outside. However, a null correlation between this item and operational indicators could reflect social desirability effects on operational indicators, or floor effects that compress variance and attenuate correlations.

Hypothesis 7: *Disagreement that ESG ratings capture performance in high-risk environments (Q17) is positively associated with higher levels of defined social outcome metrics (Q10), accessible grievance mechanisms (Q9), and expanded social engagement (Q4), suggesting that firms with greater operational maturity are more, not less, sceptical of external ESG evaluation frameworks.*

[Figure 1. Conceptual Model: Determinants of Corporate Social Purpose in Fragile Contexts]

**Methodology**

A structured synthetic survey instrument was designed targeting corporate professionals working on social impact, ESG, compliance, and supply chain management in high-risk settings. The survey captured industry, department, seniority, job title, age, gender, and geographic engagement with high-risk regions followed by 23 Likert-scale items (see Appendix 1) measured on a five-point scale from strongly disagree (1) to strongly agree (5), with additional don't know and not applicable options, with iterative review to ensure construct coverage while maintaining clarity for synthetic respondents (as compiled by Claude CoWork 1.1.772). Most items are positively worded, creating vulnerability to acquiescence bias but done to better align with existing data and LLM expectations of positivity.

The target synthetic population consisted of 400 full-time employees at firms with 1,000 or more global FTEs who work directly on issues of social impact in foreign locales that can be considered fragile, conflict-affected, or otherwise violent or insecure (200 US, 200 EU). Respondents were required to hold positions in CSR, sustainability, ESG reporting, compliance, ethics, supply chain, legal, or operations functions to capture variation across organizational hierarchies. The geographic scope was restricted to professionals headquartered in the United States or Western Europe. This dataset aimed to calibrate our hypotheses to conventional wisdoms, constructed to reproduce the distributional properties, correlation structures as specified in our theoretical framework, including predicted correlation clusters, built-in contradiction patterns (optimism bias at approximately 15% and regulatory ambivalence at approximately 20%), sector-specific response profiles, seniority effects, and systematic US-EU differentiation patterns derived from the regulatory divergence, political polarization, and institutional difference literatures.

Our approach to synthetic data occupies a distinct epistemological position from the increasingly common practice of using large language models to generate synthetic survey respondents (Argyle et al., 2023). Both approaches share a fundamental limitation: they can only reproduce patterns that are already encoded, whether in researcher-specified parameters or in LLM training corpora. The critical difference is transparency. In researcher-constructed synthetic data, the conventional wisdom that the data encode is fully specified and therefore falsifiable: every correlation structure, distributional assumption, and demographic parameter is documented, and the resulting baseline can be tested against field data to identify where professional realities diverge from scholarly expectations. LLM-generated synthetic data encode conventional wisdom implicitly, through training corpora, making it difficult to determine what the data 'know' and therefore what departures from those data would mean. Emerging literature underscores this distinction. Recently, we (redacted) tested five leading LLMs against a real human survey of 420 Silicon Valley professionals, finding that all models produced plausible and internally coherent synthetic responses that failed to capture the counterintuitive findings that constituted the human survey's most valuable insights. LLMs can simulate demographic-level opinion patterns, but synthetic replacements systematically diverge from human data on precisely the dimensions that make survey research analytically productive (Bisbee et al. 2024). Our application of synthetic data as a 'conventional wisdom baseline' rather than a substitute for empirical evidence applies: patterns are bounded by existing theoretical literature, and the true contribution will only become apparent where field data diverge from these expectations (Bender et al., 2021; Davidson & Karell, 2025).

Several parameters require specification. The seven hypothesis-level correlation clusters were designed into the data generation process: items predicted to correlate positively or negatively within each hypothesis were assigned corresponding correlation structures. Optimism bias and regulatory ambivalence were designed at specified baseline rates. However, several patterns were not directly specified: the Q14-Q8 null correlation within the US sample (the 'quiet CSR' mechanism failure) emerged from the interaction of separately specified US response distributions on these items rather than being directly programmed. Similarly, the Q14-Q24 cross-cutting correlation ($r = -.47$) arose from opposing regional distributions. The optimism bias rate (27.3% vs. the designed 15%) and its concentration among mid-level managers were also emergent properties. We distinguish emergent patterns from designed, as designed patterns confirm that the analytical framework can recover known signals, while emergent patterns suggest further investigation with field data.

This approach serves three functions. First, it provides a test of whether our analytical strategy can detect theorized patterns as a form of analytical validation for survey-based social science. Second, it allows us to calibrate effect size expectations and statistical power of $n = 400$ with a 50/50 US-EU split. Third, it surfaces potential problems: items that fail to differentiate as predicted, hypothesized correlations that are difficult to sustain even in constructed data, or contradictions in the theoretical framework itself. The synthetic results should therefore be read as a demonstration that the instrument can, in principle, capture the phenomena it was designed to measure. Certain features, such as don't know and not applicable rates, were calibrated as part of the data generation process and their primary function is to test the analytical procedures for handling missing data. For IB research in particular, where large-scale cross-national surveys of professionals in fragile contexts face substantial access, cost, and safety barriers, a validated conventional wisdom baseline allows researchers to identify where to focus analytical attention before committing to field deployment.

Therefore, we developed this approach to better tease out core conventional wisdoms. Our 23 Likert items map onto the seven hypotheses as follows. Strategic integration (H1) is captured by Q2 (strategy integration), Q6 (leadership support), Q10 (defined metrics), Q19 (supply chain-sustainability team engagement), and Q24 (CSDDD impact). Regulatory dual effect (H2) draws on Q4 (expanded engagement), Q12 (regulatory driver), and Q13 (regulatory constraint on flexibility). Political polarization (H3) centres on Q14 (polarization as obstacle) and Q8 (unreported activities, as a proxy for 'quiet CSR'). Sector-specific effects (H4) are tested through Q9 (grievance mechanisms),

Q15 (exit criteria), Q16 (security information flow), and Q22 (complicity awareness). Supply chain integration (H5) examines Q19 correlations with Q2, Q10, and Q9 within FMCG and agribusiness sectors. Subsidiary autonomy (H6) tests Q18 (local autonomy) against Q2, Q16, and Q8. ESG ratings validity (H7) examines whether Q17 (scepticism about ESG frameworks) correlates negatively with operational maturity indicators Q10, Q9, and Q4. Don't know and not applicable responses, which ranged from 1.8% (Q8) to 4.5% (Q14), were treated as missing and excluded on a pairwise basis to improve analytical power. We map these 23 items onto seven theoretical constructs, but we have not yet provided psychometric evidence that these constructs are empirically coherent (e.g., through confirmatory factor analysis). The synthetic data make factor analysis awkward, as factor structure was part of the data generation rather than an empirical discovery. Some constructs span different content domains (e.g., "strategic integration" encompasses strategy, leadership, metrics, supply chains, and CSDDD impact), and the within-construct correlations observed in the synthetic data (e.g., Q2-Q6 $r = .256$) are moderate rather than strong.

The analytical approach proceeds in four stages. First, reporting descriptive statistics for all 23 items, disaggregated by region. Second, for the five directional hypotheses predicting US-EU differences (H1, H3, and elements of H2 and H4), employing Welch's independent-samples t-tests, which do not assume equal variances across groups. Exact p-values reported alongside Cohen's d as the primary effect size metric, following the recommendation that significance testing without effect size reporting obscures practical importance (Cumming, 2014). Third, for relational hypotheses predicting within-sample correlations (H2, H5, H6, H7), Pearson's r with exact p-values are computed. Pearson's r assumes interval-level measurement, which is debatable for five-point Likert scales; Spearman's rho or polychoric correlations would provide useful robustness checks. Fourth, for sector effects (H4), one-way ANOVAs across five macro-sector groupings (extractives/energy, technology/telecom, financial services, FMCG/retail/agriculture, and defence/aerospace) with $\eta^2$ as the effect size measure, supplemented by pairwise t-tests for the specific extractives-versus-others comparisons predicted in H4.

The analysis does not apply Bonferroni or other familywise error corrections as several hypotheses make directional predictions derived from theory, and blanket correction in confirmatory frameworks is recognised as unnecessarily conservative (Rubin, 2021). Second, at $n = 200$ per group, corrected thresholds risk Type II errors on theoretically meaningful effects. Instead the full distribution of exact p-values and effect sizes are reported. Applying a Bonferroni correction would raise the significance threshold and alter the significance status of several correlations (H5, H6, H7), where the predictions are tentative and the number of tested correlations is substantial. The directional hypotheses (H1, H3), where effects are large and p-values are well below .001, would remain significant under any reasonable correction. All data and analyses were conducted in Claude CoWork v. 1.1.822 in Python 3 using scipy.stats, with pairwise deletion for missing data.

## Findings

Table 1 presents the sample composition, and Table 2 reports descriptive statistics for all 23 Likert items. Across the sample, means ranged from 2.79 (Q24: CSDDD impact) to 3.74 (Q6: leadership support), with standard deviations between 0.93 and 1.46. DK/NA rates were low, ranging from 1.8% to 4.5%. 19 items showed statistically significant US-EU differences ($p < .05$), indicating that the instrument successfully captures the regulatory and political divergences embedded in the theoretical framework. Table 3 summarises the hypothesis test results.

[Table 1: Sample Characteristics (N = 400)]

[Table 2: Descriptive Statistics and US-EU Comparisons for All Likert Items]

H1: Regulatory divergence and strategic integration

The first hypothesis predicted that EU professionals would report higher levels of strategic integration of social impact initiatives than their US counterparts. This prediction was supported across all five constituent items. The largest effect appeared on Q24, where EU respondents reported substantially higher agreement that the CSDDD had improved their social activities (EU M = 4.01, US M = 1.58, d = -2.97, $p < .001$). This suggests that the CSDDD has produced an experiential divide, with EU professionals affirming its transformative impact and US professionals registering low levels of agreement. The remaining items showed more moderate EU advantages: defined social outcome metrics (Q10, d = -0.60), supply chain-sustainability team engagement (Q19, d = -0.57), strategy integration (Q2, d = -0.51), and leadership support (Q6, d = -0.29). The gradient of effect sizes is informative: the strongest effects cluster around items with the most direct regulatory nexus (Q24, Q10, Q19), while the weakest (leadership support) capture organisational variables that are arguably less proximate to regulatory compliance.

H2: The regulatory dual effect

Hypothesis 2 predicted that regulatory pressure would be positively associated with both reported social engagement and perceived constraints on flexibility, a dual effect in which mandatory due diligence regimes simultaneously compel action and create bureaucratic burden. The directional component was supported: EU respondents reported significantly higher regulatory-driven engagement (Q12, d = -0.74, the second-largest non-Q24 effect in the survey) and expanded social engagement over the past three years (Q4, d = -0.48). However, the within-region correlation between Q12 and Q13 was not significant in either sample (EU r = -.11, p = .148; US r = .09, p = .240), suggesting that the regulatory ambivalence pattern operates at the individual respondent level rather than as a linear item-to-item relationship. This is confirmed by the contradiction analysis: 23.4% of respondents simultaneously agreed with both Q12 ('regulations drive engagement') and Q13 ('regulations reduce flexibility'), with this pattern concentrated among EU respondents (27.0%) relative to US respondents (19.7%).

H3: Political polarization and the quiet CSR puzzle

US professionals reported significantly higher agreement that political polarization had made it harder to pursue social initiatives in high-risk environments (Q14: US M = 4.16, EU M = 2.87, d = 1.38, $p < .001$). This is the largest US-EU gap in the survey apart from Q24, and its effect size places it above conventional thresholds for a 'large' effect. For context, fewer than 8% of US respondents disagreed or strongly disagreed with Q14, compared to over 40% of EU respondents. The asymmetry reflects a qualitatively different professional experience of the political environment surrounding corporate social purpose. The hypothesised mechanism, however, did not materialise. Expectation is that US professionals who perceived higher polarization (Q14) would also report more unreported social activities (Q8), the 'quiet CSR' phenomenon whereby firms continue social programming while reducing public communication. The within-US correlation between Q14 and Q8 was essentially zero (r = -.01, p = .90). This null correlation was not directly designed into the synthetic data but emerged from the interaction of separately specified response distributions, making it a more interesting finding than a simple confirmatory result.

US respondents did offer higher unreported activities than EU respondents (Q8: US M = 3.61, EU M = 3.40, d = 0.22, p = .032), and US respondents also reported higher agreement that CSR is more relevant to marketing than operations (Q21: US M = 3.52, EU M = 2.78, d = 0.73, p < .001). The quiet CSR phenomenon appears to operate through channels other than the direct perception of political hostility, perhaps through institutional norms, legal risk aversion, or organisational culture rather than individual-level political anxiety. Q8 merges genuine greenhushing, routine organisational slack, and structural disconnection between operations and reporting functions.

### H4: Extractive sector distinctiveness

Hypothesis 4 predicted that professionals in extractive industries would report higher operational preparedness for conflict escalation alongside greater willingness to acknowledge that their firm's presence can worsen local dynamics. Both components were supported. One-way ANOVAs across the five macro-sector groupings revealed significant between-sector variation on Q15 (exit criteria, F = 6.52, p < .001, $\eta^2$ = .09), Q16 (security information flow, F = 6.93, p < .001, $\eta^2$ = .10), and Q22 (complicity awareness, F = 4.80, p < .001, $\eta^2$ = .07). In each case, extractives professionals reported the highest group mean: Q15 (M = 3.80 vs. 3.04-3.55 for other sectors), Q16 (M = 3.87 vs. 2.76-3.45), and Q22 (M = 3.58 vs. 2.78-3.13). The extractives-versus-all-others comparison on Q22 yielded a medium effect (d = 0.51, p < .001).

This pattern is consistent with the literature on corporate complicity in conflict settings (Kim & Davis, 2016): firms with the most direct physical presence in fragile environments develop both the operational infrastructure to manage instability and the reflexive awareness that their presence carries risks for host communities. It is notable that technology and financial services professionals reported the lowest means on Q15 and Q16, suggesting that firms with more mediated (and perhaps more easily disengaged) modes of engagement in fragile contexts invest less in operational preparedness for escalation. The defence/aerospace sector (n = 20) also reported elevated scores on these items. Caution is warranted given the small subsample, though the pattern is consistent with the physical-presence interpretation.

### H5 and H6: Supply chain integration and subsidiary autonomy

Hypothesis 5 was partially supported. Within FMCG and agribusiness sectors (n = 70), supply chain-sustainability team engagement (Q19) correlated significantly with the presence of accessible grievance mechanisms (Q9, r = .40, p < .001), but not with broader strategic integration (Q2, r = .09, p = .45) or defined social outcome metrics (Q10, r = .16, p = .20). This suggests that supply chain-sustainability integration in these sectors translates into specific operational mechanisms like downward accountability channels but does not scale to the strategic level. The finding is consistent with the observation that supply chain due diligence tends to produce compliance-oriented practices before it reshapes corporate strategy (Torelli et al., 2025).

Hypothesis 6 was fully supported: all three predicted subsidiary autonomy relationships were confirmed. Local autonomy (Q18) correlated negatively with strategic integration (Q2, r = -.19, p < .001) and with the timeliness of security information reaching headquarters (Q16, r = -.18, p < .001), but positively with unreported social activities (Q8, r = .17, p < .001). The HQ-subsidiary tension is real: greater local autonomy purchases operational flexibility at the cost of strategic coherence and informational transparency. This negative correlation may reflect a selection effect (Ambos et al., 2010): firms may grant more autonomy to subsidiaries in contexts where centralised strategy is impractical, making the association an artefact of operational context rather than evidence of organisational tension. US respondents reported significantly higher subsidiary autonomy than EU respondents (Q18, d = 0.51), a difference that aligns with the more centralised, compliance-driven governance structures that the CSDDD and its national predecessors impose on EU firms.

H7: The ESG ratings null finding

Hypothesis 7, that scepticism about ESG ratings' validity in high-risk environments would be positively associated with operational maturity, was not supported. None of the three predicted negative correlations between Q17 and operational indicators (Q10, Q9, Q4) approached significance ($r$ values ranged from -.05 to .08, all $p > .14$). When comparing respondents who disagreed with Q17 ('ESG sceptics', scores $\leq 2$) against those who agreed ('ESG believers', scores $\geq 4$), the two groups were statistically indistinguishable on defined metrics (Q10), grievance mechanisms (Q9), and expanded engagement (Q4). This null finding was not a designed parameter of the synthetic data; the correlation between Q17 and operational items was left unspecified in the data generation process, allowing it to emerge (or not) from the broader correlation structure.

The null finding is theoretically consequential. It suggests that ESG ratings scepticism may be inverse to operational sophistication: professionals who recognise the limitations of ESG frameworks are not those who built superior internal systems to compensate. If anything, scepticism and maturity appear to be independent dimensions, complicating the narrative that operational experience breeds institutional critique. However, the null correlation could reflect poor construct validity of Q17, social desirability effects on the operational indicators, or floor effects on Q17 ($M = 2.95$, close to the midpoint) that compress variance and attenuate correlations. The leap from 'no correlation between one survey item and three others' to broader claims about ESG measurement frameworks requires evidence that this instrument does not yet provide.

[Table 3: Summary of Hypothesis Tests]

[Table 4: Key Correlation Pairs by Theoretical Cluster]

### *Emergent patterns: contradictions and cross-cutting effects*

Two designed contradiction patterns and one emergent cross-cutting effect merit discussion. The optimism bias pattern (respondents reporting high leadership support ($Q6 \geq 4$) and low willingness to sacrifice profitability ($Q5 \leq 2$)) appeared at 27.3% of the sample, substantially above the designed baseline of 15%. The pattern was more prevalent among US respondents (31.7%) than EU respondents (23.0%), and counterintuitively, more common among mid-level professionals (23.8%) than senior leaders (16.4%). Both were emergent properties of the simulation, suggesting that the interaction of separately specified response distributions on Q5 and Q6 produces amplification effects. This reversal of the expected C-suite optimism pattern may reflect the greater exposure of mid-level managers to the gap between stated commitment and operational reality, a form of 'decoupled optimism' in which rhetorical support from above is perceived but not matched by resource allocation visible at the implementation level.

The regulatory ambivalence pattern (agreement that regulations drive engagement ($Q12 \geq 4$) and reduce flexibility ($Q13 \geq 4$)) appeared at 23.4%, close to the designed 20%. Its concentration among EU respondents (27.0% vs. 19.7% for US) is consistent with the interpretation that mandatory due diligence creates a lived paradox: professionals in these settings simultaneously credit regulation for expanding their social mandate and fault it for constraining their ability to respond authentically to local conditions. The strongest bivariate correlation in the entire matrix, between Q14 (political polarization as obstacle) and Q24 (CSDDD impact), was negative and substantial ($r = -.47$, $p < .001$).

This reveals that respondents who report high political polarization report low CSDDD impact, and vice versa. This US anti-ESG/EU pro-regulation axis may represent the single most consequential fault line in contemporary corporate social purpose, cleaving the professional experience of social impact work along a transatlantic regulatory and political divide.

**Discussion**

These synthetic findings offer a foundation for rethinking how corporate social purpose conventional wisdoms may manifest in contexts of crisis and institutional fragility. Next, I discuss four themes for advancement in business, society, and conflict.

The most robust pattern in the synthetic data is the transatlantic divergence on strategic integration (H1), where EU professionals reported higher scores on all five constituent items, with effects ranging from moderate (Q6, leadership support, $d = 0.29$) to extraordinary (Q24, CSDDD impact, $d = 2.97$). The gradient of effect sizes is theoretically instructive. The weakest effect appears on the item most distal from regulatory compulsion (leadership support), while the strongest cluster around items with direct regulatory nexus: defined metrics, supply chain coordination, and the CSDDD itself. This pattern may extend the political CSR framework (Scherer et al. 2016), as the synthetic data suggest that mandatory due diligence legislation may reconstitute the organisational logics through which social impact is conceived, measured, and integrated into corporate strategy.

The institutional work perspective (Lawrence et al., 2009; 2011) offers useful analytical resources. Mandatory due diligence may trigger active processes of institutional creation within organisations: professionals develop new measurement systems, coordination routines, and reporting architectures that reshape how the organisation understands its social role. This goes beyond coercive isomorphism (DiMaggio & Powell, 1983), in which organisations adopt similar structures under regulatory pressure, because the content of what is being adopted (strategic integration of social impact) requires substantive organisational change rather than merely formal compliance. For IB scholarship, the finding carries a specific implication: home-country regulatory institutions may shape not only compliance behaviour but the organizational logics through which MNCs understand their social purposes in host countries. The gradient of effect sizes, strongest on items with direct regulatory nexus and weakest on items distal from regulatory compulsion, is consistent with an institutional distance interpretation in which mandatory due diligence narrows the gap between home-country expectations and host-country practice by restructuring the headquarters-level governance apparatus.

How does this dynamic differ from Greenwood's (2011) institutional complexity as in situations where organisations face incompatible prescriptions from multiple institutional logics. The regulatory dual effect (H2) appears useful: the CSDDD simultaneously compels engagement and constrains flexibility, producing contradictory demands. But the tension may operate within a single regulatory logic. The 23.4% of respondents who simultaneously affirmed that regulation drives engagement and constrains flexibility experienced the internal contradictions of a regulatory mandate that demands both standardised compliance and context-sensitive implementation. I propose the term constitutive tension to capture this specific dynamic: a condition in which regulation does not supplement existing organisational practice but restructures the terms on which practice is understood, generating new capacities and new frustrations in the same institutional movement. The concept aligns with institutional complexity but differs in locating the tension within the logic of mandatory due diligence, and with paradox theory's performing paradoxes (Smith & Lewis, 2011) but emphasises the regulatory origin of the contradiction rather than treating it as an inherent feature of organisational life. Recent work on the CSDDD as a "paradigm shift" in global value chain governance (Wilhelm, 2024) supports this interpretation, as does Sinnig and Zetzsche's (2024) analysis of the directive's embedding mechanisms.

The constitutive tension pattern aligns with conventional wisdom in the political CSR literature: that mandatory regulation reshapes organisational logics, not merely compliance behaviour. The synthetic data confirm this pattern, yet it is unclear whether the magnitude and structure of the effect replicate, or whether the lived experience of practitioners under mandatory due diligence produces patterns that the literature has not anticipated (Miklian et al., 2026). The theoretical implication is that regulation should be analysed not through the lens of compliance costs and incentive structures, as neoclassical treatments suggest, but as a constitutive institutional force that reshapes how organisations understand their own social purposes. The Q24 binary split, where EU respondents cluster at the top of the scale and US respondents at the floor, is consistent with two professional populations operating within different institutional realities, where the very meaning of corporate social purpose in fragile contexts has been reconfigured by the regulatory environment.

The political polarization findings (H3) generated the survey's most striking single result (Q14, d = 1.38) alongside its most instructive mechanism failure. US professionals overwhelmingly report that political polarization has made it harder to pursue social initiatives in high-risk environments. Fewer than 8% of US respondents disagreed with Q14, compared to over 40% of EU respondents. This asymmetry is consistent with anti-ESG backlash (Tang et al. 2024; Curtis 2025), as in which more than 480 anti-ESG bills had been introduced across US state legislatures by early 2026.

Yet the predicted mechanism did not hold in the synthetic data. Perceived political hostility (Q14) was hypothesized to predict unreported social activities (Q8), producing the "quiet CSR" pattern in which firms continue social programming while suppressing public communication. The within-US Q14-Q8 correlation was zero (r = -.01). US respondents did report higher unreported activities than EU respondents (Q8, d = 0.22), and higher agreement that CSR is more relevant to marketing than operations (Q21, d = 0.73). But these patterns are not connected to polarization perceptions at the individual level. The quiet CSR phenomenon, if it exists as a strategic response to political hostility, operates through channels other than direct political anxiety.

The Q14 directional finding delivers conventional wisdom that US professionals perceive political polarization as an obstacle to social engagement — just what the anti-ESG literature would expect. The null Q14-Q8 mechanism, by contrast, represents the counterintuitive pattern that Miklian et al. (2026) found LLM-generated synthetic data systematically unable to produce; that it emerged from researcher-constructed data rather than a designed parameter suggests the instrument has properties worth investigating. This distinction is methodologically consequential: researcher-constructed synthetic data, precisely because they make their conventional wisdom explicit, can generate emergent patterns that were not designed into the data. The Q14-Q8 null emerged from the interaction of separately specified response distributions, demonstrating that the conventional wisdom baseline approach has diagnostic properties beyond simple confirmation. The emerging literature on greenhushing, deliberate underreporting of genuine sustainability activity to avoid reputational exposure (Montgomery et al. 2024), has framed strategic silence as a rational response to institutional complexity. The synthetic data complicate this framing, or at least suggest a refinement. If greenhushing were driven by perceived political threat, we should observe a correlation between Q14 and Q8 among US respondents.

Its absence suggests that corporate silence on social impact may be better understood through institutional lenses. Drawing on Jamali et al's (2017) concept of selective decoupling, unreported social activity in high-risk US contexts may reflect not a calculated response to backlash but an institutionalised communication norm in which social impact work in fragile settings was never integrated into public-facing corporate narratives in the first place. The anti-ESG environment may have reinforced this norm, but it did not create it. The Q21 finding (US respondents viewing CSR as more marketing than operations, d = 0.73) is consistent with this reading: in organisations where social impact is coded as a communications function, unreported activities are not "hidden" so much as structurally invisible to the reporting apparatus. If corporate social engagement in fragile contexts is structurally decoupled from corporate communications in US firms, then the informational basis on which scholars, investors, and regulators assess that engagement is weak. The calls for greater transparency that pervade the ESG reform literature (Berg et al. 2022) may be addressing the wrong problem. The issue may not be that firms hide their activities but that the organisational architectures through which social impact is managed and communicated were built for different purposes.

Hypothesis 7's null result, that ESG ratings scepticism (Q17) bore no relationship to operational maturity indicators (Q10, Q9, Q4), is perhaps the most consequential finding for the field's future direction. Professionals who recognised the limitations of ESG frameworks in high-risk environments may be those whose organisations had built superior internal measurement and accountability systems, a compensatory logic in which institutional critique drives operational investment. The synthetic data do not support this logic. Sceptics and believers are statistically indistinguishable on defined metrics, grievance mechanisms, and expanded engagement.

This deserves interpretation, with the caveat that null findings in synthetic data occupy an ambiguous epistemic position: the absence of a correlation may reflect genuine independence of the constructs, or it may indicate that the data generation process did not specify a relationship that exists in practice. It does not mean that ESG frameworks work well in fragile contexts. The Q17 mean (2.95) suggests widespread scepticism, and the EU advantage on Q17 (d = 0.29) suggests that even mandatory reporting regimes do not resolve practitioners' doubts about whether ratings capture what matters. Scepticism and operational sophistication may be independent dimensions of professional experience. The expected convergence, where experience breeds critique and critique drives improvement, does not appear in the synthetic data, in support of Berg et al. (2022).

Where their analysis documents divergence across rating agencies at the firm level, these synthetic data point toward a more fundamental problem at the context level: ESG measurement frameworks may be porly suited to environments characterised by instability, information asymmetry, and rapid change. The fragile-context practitioner's scepticism would not be about which rating agency to trust but about whether the measurement paradigm itself can accommodate the phenomena it claims to capture. For corporate governance scholarship, this points toward a research agenda on context-sensitive measurement that moves beyond disclosure-based metrics toward indicators grounded in operational realities of high-risk environments, what can be called corporate social irresponsibility in foreign subsidiary performance (Nuruzzaman et al. 2024).

The extractives sector findings (H4) open a fourth line of theoretical development that cuts across the preceding three. Extractives professionals reported the highest scores on exit criteria (Q15, M = 3.80), security information flow (Q16, M = 3.87), and complicity awareness (Q22, M = 3.58), with sector effects explaining 7-10% of variance on these items ($\eta^2 = .07-.10$). Technology and financial services professionals reported the lowest scores on Q15 and Q16, while financial services reported the lowest complicity awareness (Q22, M = 2.78). What distinguishes these sectors is not their level of engagement with fragile contexts but their mode of presence: extractive firms maintain direct physical operations in fragile environments, while technology and financial services engage through intermediated, often digital, channels that are easier to scale back or exit. This variation structures how firms understand their own role. Extractives professionals develop both the operational infrastructure to manage instability (exit criteria, information systems) and the reflexive awareness that their presence carries risks for host communities. This dual pattern is absent in sectors with more mediated engagement. The finding resonates with Ganson and Wennmann's (2016) argument that pragmatic engagement with conflict requires confronting uncomfortable truths about corporate complicity, and with Kim and Davis's (2016) demonstration that physical supply chain presence in conflict zones produces qualitatively different accountability dynamics. This finding extends the IB literature on mode of entry and commitment depth into the social domain: the deeper a firm's physical commitment to a fragile host environment, the more developed both its operational infrastructure and its capacity for reflexive self-assessment.

The sector differentiation pattern is partly conventional wisdom (business and peace literature has long argued that mode of engagement matters) and partly novel in its operationalisation. Whether the dual pattern of preparedness and self-critique replicates in field data, or whether real extractive sector professionals display different configurations, will test the degree to which a firm's self-understanding of its social role in fragile contexts is shaped by the physical, operational, and relational depth of its engagement, including the capacity for self-criticism. The defence and aerospace sector respondents reinforce this interpretation, showing elevated scores on operational preparedness items that parallel the extractives pattern. The partially supported supply chain integration hypothesis (H5) adds texture: in FMCG sectors, supply chain and sustainability team coordination (Q19) predicted grievance mechanisms (Q9, $r = .40$) but not broader strategic integration (Q2) or defined metrics (Q10). Supply chain linkages generate specific accountability mechanisms, a form of operational reflexivity, without scaling to the strategic level. The subsidiary autonomy findings (H6) complete the picture: greater local autonomy (Q18) trades strategic coherence and informational transparency for operational flexibility, a trade-off that is more pronounced in US firms (Q18, $d = 0.51$, US higher) than in EU firms, consistent with the more centralised governance structures that mandatory due diligence imposes. For IB scholarship on headquarters-subsidiary relations, this offers rare cross-national survey evidence that regulatory context shapes the integration-responsiveness balance in the social governance of MNC operations in high-risk environments, a context the IB subsidiary literature has largely overlooked.

**Limitations and Forward Research**

As mentioned, the primary limitation of this paper is that results are based on synthetic data. The patterns identified are consistent with our theoretical framework because the data were constructed to contain those designed patterns or to allow unspecified patterns to emerge. The proposed theoretical contributions are propositions for empirical testing. Researcher-constructed synthetic data avoids some LLM pitfalls since we can control the parameter space, but a core epistemological concern applies: synthetic data can confirm what we already expect while remaining silent on what we do not yet know. Therefore, this is a synthesis of existing knowledge into a digestible and comparable format. Unlike LLM-generated synthetic surveys, this approach makes the conventional wisdom it encodes fully transparent, documented, and therefore falsifiable against field data. Future study will determine which of the patterns identified here also reflect professional reality, and which challenge them.

Three other limitations warrant discussion. The first concerns causal inference. Our relational hypotheses (H2, H5, H6, H7) predict associations between survey items, but the cross-sectional design cannot establish directionality. When reporting that subsidiary autonomy correlates negatively with strategic integration (H6), it is unclear whether decentralised governance produces strategic fragmentation, whether weak strategic integration leads firms to delegate authority downward, or whether both are driven by an unmeasured third variable such as organisational size, maturity, or industry norms. The same ambiguity applies to the regulatory dual effect (H2): EU professionals report higher engagement and higher perceived constraint, but whether mandatory due diligence causes both outcomes or whether pre-existing organisational orientations sort firms into compliance profiles that produce this pattern remains an open question. Longitudinal designs tracking the same firms across regulatory transitions, for instance before and after CSDDD transposition into national law, would permit stronger causal claims. Panel surveys administered at two or three time points during the directive's implementation phase (2027 onward) represent a natural next step. The description of the US-EU regulatory divergence requires qualification. A natural experiment requires that assignment to treatment and control conditions is plausibly exogenous. Firms' headquarters locations reflect pre-existing institutional, cultural, and strategic orientations that also predict social engagement patterns. EU-US differences could reflect regulatory effects, but they could equally reflect longstanding differences in stakeholder capitalism traditions, labour market institutions, or corporate governance cultures.

The second limitation is the gap between perceptions and practice. Our data capture conventional wisdoms on what professionals perceive about their firms' social engagement in fragile contexts. They do not capture what firms do. The decoupling literature has documented persistent misalignment between corporate policies and operational realities, particularly in CSR domains where reputational incentives reward symbolic commitment over substantive action. Social desirability bias may inflate reported levels of strategic integration, leadership support, and grievance mechanism accessibility. It may also depress complicity awareness (Q22) and unreported activities (Q8), where candour carries professional risk. There is no independent verification of how this data corresponds to observable outcomes. Future research could triangulate survey findings with disclosed reporting data, third-party assessments, and where possible, community-level evaluations of corporate behaviour in specific fragile settings.

The third limitation is positionality. The synthetic survey focused on corporate headquarters, not the communities in fragile settings whose welfare these corporate initiatives purport to serve. The perceptions of a sustainability director in Frankfurt or Chicago about the effectiveness of their firm's grievance mechanisms in the Democratic Republic of Congo tell us something about organisational self-understanding but nothing about whether those mechanisms function as intended for the people who need them. This limitation helps to sharpen the findings: HQ-level perceptions shape resource allocation, strategic priorities, and compliance decisions, and the existing corpus of information leans heavily towards this information, enhancing the ability to deliver accurate synthetic conventional wisdoms; this data does not exist at the ground level.

Taken together, this exercise suggests that corporate social purpose in fragile and conflict-affected settings is not a unitary phenomenon amenable to a single theoretical lens. It is shaped by regulatory environments that may constitute rather than constrain, by political economies that produce silence as much as speech, by measurement paradigms that may be poorly suited to the contexts they claim to evaluate, and by modes of presence that structure the very capacity for organisational self-understanding. The field's theoretical architecture, built on assumptions of voluntary action, transparent communication, and universal measurement, may require revision to accommodate these realities. For international business scholarship, these findings suggest that the study of MNCs in fragile contexts requires closer attention to how home-country regulatory and political environments reshape the organizational logics governing social engagement abroad. The IB literature has extensively theorised headquarters-subsidiary governance, institutional distance, and integration-responsiveness tensions, but these constructs have rarely been applied to corporate social engagement in the most challenging operational environments. The conventional wisdom baseline presented here makes these dynamics empirically tractable, and field validation will determine whether the patterns that theory predicts also describe the professional realities of MNC social governance in contexts of crisis and fragility. Future research should examine whether patterns identified here (a transatlantic regulatory divide, a decoupled political economy of corporate silence, the inversality of measurement scepticism and operational maturity, presence-dependent reflexivity) hold when tested with field data from the populations these synthetic data were designed to represent.

# Figures and Tables

**Figure 1. Conceptual Model: Determinants of Corporate Social Purpose in Fragile Contexts**

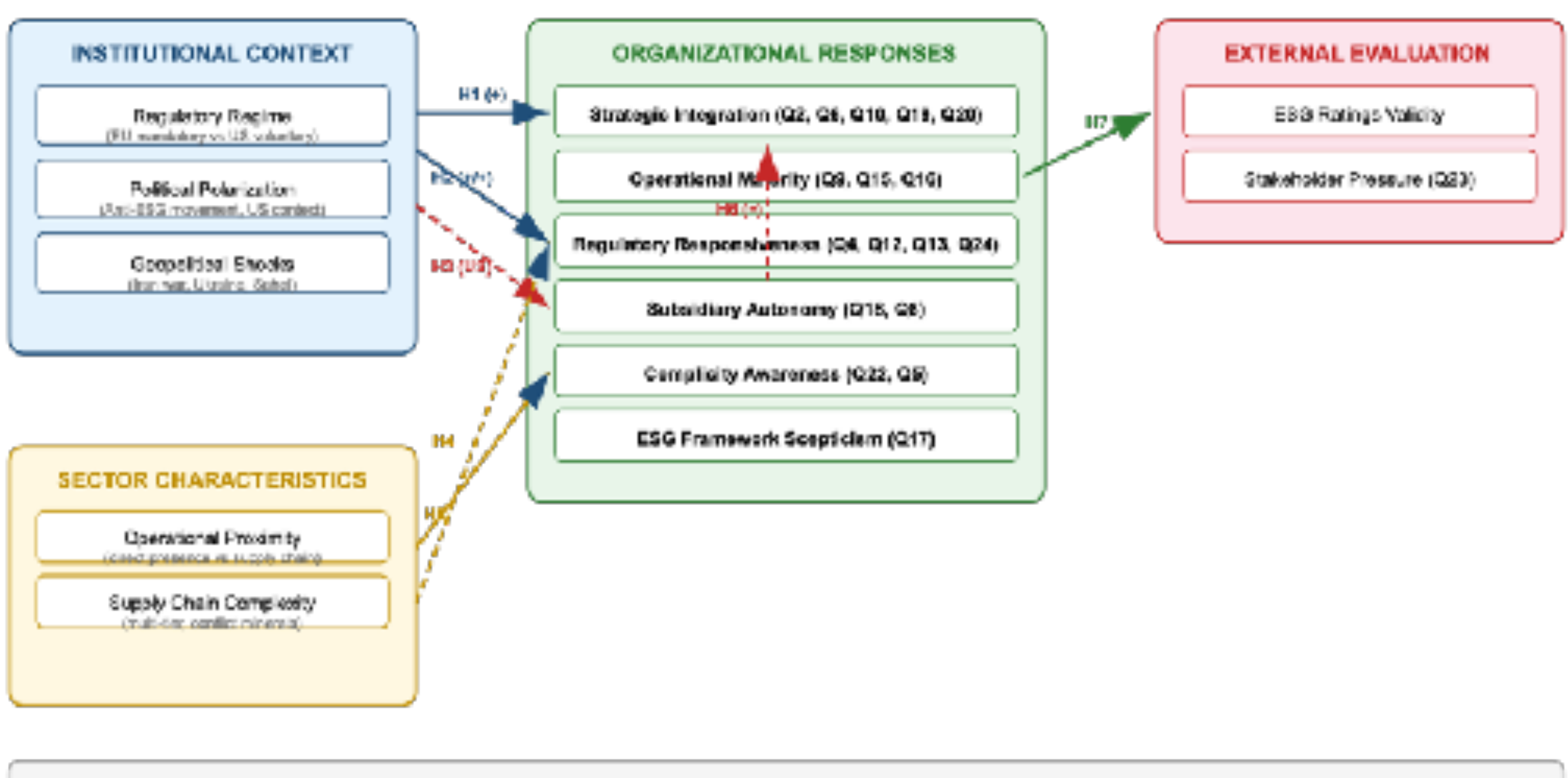

**Hypothesis Summary**

H1: EU firms report higher strategic integration than US firms (Q2, Q6, Q10, Q19)
H2: Regulatory pressure positively associated with BOTH social engagement AND perceived bureaucratic constraints (Q4, Q12, Q13)
H3: US professionals report higher political polarization impact (Q14); positively associated with unreported activities (Q8)
H4: Extractive firms: higher operational preparedness (Q9, Q15, Q16) AND higher complicity awareness (Q22)
H5: In supply chain sectors, supply chain-sustainability integration (Q19) clusters with strategic maturity (Q2, Q10, Q20)
H6: Subsidiary autonomy (Q18) negatively associated with strategic integration (Q2) and HQ info flow (Q10); positively with unreported activities (Q8)
H7: ESG ratings scepticism (Q17) positively associated with operational maturity (Q13, Q9, Q4) — mature firms are more sceptical

Table 1
**Sample Characteristics (N = 400)**

| Characteristic | Category | US (n) | EU (n) | Total |
|---|---|---|---|---|
| Sector (S1) | Extractives/Mining/Energy | 34 | 32 | 66 |
| | Financial services | 28 | 28 | 56 |
| | Technology/Software/Telecom | 28 | 28 | 56 |
| | FMCG/Retail/Agriculture | 36 | 34 | 70 |
| | Defense/Aerospace | 10 | 10 | 20 |
| | Other | 64 | 68 | 132 |
| Seniority (S3) | VP and above | 38 | 34 | 72 |
| | Director/Sr. Director | 54 | 54 | 108 |
| | Manager/Contributor | 108 | 112 | 220 |
| Recency (S8) | New (conflict-related) | 45 | 48 | 93 |
| | Ongoing | 155 | 152 | 307 |

*Note. Sector groupings aggregate S1 categories. 'Other' includes construction, pharmaceuticals, professional services, and miscellaneous. Seniority groups: VP and above includes VP (EVP/SVP) and*

*department/line-of-business heads; Manager/Contributor includes directors, senior managers, managers, and individual contributors.*

Table 2
Descriptive Statistics and US-EU Comparisons for All Likert Items

| Item | Question text (abbreviated) | All M | All SD | US M | US SD | EU M | EU SD | d | p | DK % |
|---|---|---|---|---|---|---|---|---|---|---|
| Q1 | Greater responsibility in high-risk | 3.66 | 1.01 | 3.41 | 1.04 | 3.92 | 0.94 | -0.53 | <.001 | 3.2 |
| Q2 | Social impact integrated in strategy | 3.29 | 1.09 | 3.03 | 1.06 | 3.56 | 1.06 | -0.51 | <.001 | 3.2 |
| Q3 | More operations in high-risk (3yr) | 3.72 | 0.93 | 3.70 | 0.95 | 3.74 | 0.92 | -0.04 | .686 | 3.0 |
| Q4 | Expanded social engagement (3yr) | 3.48 | 0.94 | 3.26 | 0.92 | 3.70 | 0.89 | -0.48 | <.001 | 4.0 |
| Q5 | Social resp. at expense of profit | 2.98 | 1.07 | 2.80 | 1.06 | 3.17 | 1.06 | -0.34 | <.001 | 3.8 |
| Q6 | Top leadership supports social impact | 3.74 | 1.04 | 3.59 | 1.05 | 3.89 | 1.01 | -0.29 | .005 | 4.0 |
| Q7 | Monitor competitors on CSR | 3.46 | 1.03 | 3.24 | 1.03 | 3.69 | 0.99 | -0.45 | <.001 | 3.5 |
| Q8 | Unreported social activities | 3.51 | 1.00 | 3.61 | 0.96 | 3.40 | 1.03 | 0.22 | .032 | 1.8 |
| Q9 | Accessible grievance mechanisms | 3.47 | 0.97 | 3.30 | 0.97 | 3.64 | 0.93 | -0.36 | <.001 | 2.0 |
| Q10 | Defined social outcome metrics | 3.41 | 1.15 | 3.08 | 1.15 | 3.74 | 1.06 | -0.60 | <.001 | 3.8 |
| Q12 | Regulatory requirements drive engagement | 3.48 | 1.05 | 3.11 | 0.99 | 3.84 | 0.97 | -0.74 | <.001 | 3.5 |
| Q13 | Regulations reduce flexibility | 3.34 | 1.00 | 3.44 | 0.99 | 3.25 | 1.01 | 0.19 | .058 | 3.5 |
| Q14 | Political polarization hinders social | 3.52 | 1.13 | 4.16 | 0.77 | 2.87 | 1.06 | 1.38 | <.001 | 4.5 |
| Q15 | Clear exit criteria for escalation | 3.25 | 1.09 | 3.14 | 1.08 | 3.36 | 1.10 | -0.20 | .045 | 3.0 |
| Q16 | Security info reaches HQ timely | 3.26 | 1.15 | 3.20 | 1.16 | 3.33 | 1.14 | -0.12 | .258 | 3.5 |
| Q17 | ESG ratings capture high-risk perf. | 2.95 | 1.02 | 2.81 | 1.01 | 3.10 | 1.01 | -0.29 | .004 | 3.5 |
| Q18 | Local subsidiaries have autonomy | 3.24 | 1.01 | 3.48 | 0.96 | 2.99 | 1.01 | 0.51 | <.001 | 3.0 |
| Q19 | SC and sustainability teams engage | 3.30 | 1.12 | 2.99 | 1.09 | 3.60 | 1.08 | -0.57 | <.001 | 4.0 |
| Q20 | Team engages with senior mgmt | 3.30 | 1.08 | 3.19 | 1.10 | 3.41 | 1.06 | -0.21 | .042 | 4.2 |
| Q21 | CSR more marketing than operations | 3.15 | 1.07 | 3.52 | 1.02 | 2.78 | 1.02 | 0.73 | <.001 | 3.2 |
| Q22 | Presence can worsen conflict dynamics | 3.14 | 1.05 | 2.98 | 1.05 | 3.30 | 1.02 | -0.31 | .003 | 3.2 |
| Q23 | Outside actors make tangible difference | 3.50 | 0.98 | 3.39 | 0.98 | 3.60 | 0.97 | -0.21 | .039 | 3.8 |
| Q24 | CSDDD changed activities for better | 2.79 | 1.46 | 1.58 | 0.82 | 4.01 | 0.92 | -2.97 | <.001 | 2.5 |

*Note. M = mean; SD = standard deviation; d = Cohen's d (negative values indicate EU > US); DK% = percentage of don't know/not applicable responses. Welch's t-test (unequal variances).*

Table 3
Summary of Hypothesis Tests

| | Hypothesis | Test | Key results | Status |
|---|---|---|---|---|
| H1 | EU higher strategic integration (Q2, Q6, Q10, Q19, Q24) | Welch's t | 5/5 items EU > US ($p < .05$); d = 0.29-2.97 | Supported |
| H2 | Regulatory pressure -> engagement + constraints | t-test; r | EU higher Q12 (d = 0.74), Q4 (d = 0.48); Q12-Q13 r = -.04 ns | Supported |
| H3 | US higher polarization (Q14); quiet CSR (Q14->Q8) | t-test; r | Q14 d = 1.38; US Q14-Q8 r = -.01 ns | Supported (partial mechanism) |
| H4 | Extractives: higher ops preparedness + complicity | ANOVA; t | Q15 F = 6.52, eta-sq = .09; Q22 d = 0.51 (Ext. vs others) | Supported |
| H5 | FMCG: SC-sustainability integration -> strategic | r (FMCG) | Q19-Q9 r = .40*; Q19-Q2 r = .09 ns; Q19-Q10 r = .16 ns | Partially supported |
| H6 | Subsidiary autonomy inverse relationships | r | Q18-Q2 r = -.19*; Q18-Q16 r = -.18*; Q18-Q8 r = .17* | Supported |
| H7 | ESG scepticism -> operational maturity | r | Q17-Q10 r = .08 ns; Q17-Q9 r = -.05 ns; Q17-Q4 r = .01 ns | Not supported |

*Note. d = Cohen's d; r = Pearson's r; eta-sq = eta-squared. * $p < .05$, ** $p < .01$, *** $p < .001$. Welch's t-tests for group comparisons; one-way ANOVA for sector effects.*

Table 4
Key Correlation Pairs by Theoretical Cluster

| Cluster | Item A | Item B | r | p | Validation |
|---|---|---|---|---|---|
| Strategic Integration | Q2 (Strategy) | Q10 (Metrics) | .334 | <.001 | Validated |
| | Q2 (Strategy) | Q6 (Leadership) | .256 | <.001 | Validated |
| | Q2 (Strategy) | Q20 (Sr. mgmt) | .279 | <.001 | Validated |
| Regulatory | Q12 (Reg. drive) | Q24 (CSDDD) | .345 | <.001 | Validated |
| | Q12 (Reg. drive) | Q13 (Flexibility) | -.037 | .478 | Not validated |
| Operational | Q9 (Grievance) | Q19 (SC teams) | .330 | <.001 | Validated |
| | Q9 (Grievance) | Q10 (Metrics) | .254 | <.001 | Validated |
| | Q9 (Grievance) | Q16 (Info flow) | .230 | <.001 | Validated |
| Autonomy | Q18 (Autonomy) | Q2 (Strategy) | -.186 | <.001 | Validated |
| | Q18 (Autonomy) | Q16 (Info flow) | -.182 | <.001 | Validated |
| | Q18 (Autonomy) | Q8 (Unreported) | .172 | <.001 | Validated |
| Marketing-Ops | Q21 (Marketing) | Q2 (Strategy) | -.238 | <.001 | Validated |
| | Q21 (Marketing) | Q10 (Metrics) | -.206 | <.001 | Validated |
| Cross-cutting | Q14 (Polarization) | Q24 (CSDDD) | -.467 | <.001 | Strongest r |

*Note. Pearson's r, pairwise deletion. N ranges from 368 to 381 depending on item-pair DK/NA rates. Bold cluster names indicate first entry for that cluster; subsequent rows within the same cluster are sub-pairs. The strongest correlation in the full matrix (Q14-Q24, r = -.467) is cross-cutting between the regulatory and polarization constructs.*

**AI Statement:** Claude CoWork 1.1.772 was used as the primary tool to develop the synthetic data, tables, and figures, alongside offering suggestions for fit and findings precision.

**Appendix: Survey**

*(Note all questions are designed for Likert-scale responses: Strongly disagree, Disagree, Neither agree nor disagree, Agree, Strongly agree, Don't know, Not applicable)*

1. My company has a greater responsibility to contribute to society in high-risk environments than in more stable regions of operation.

2. My company's social improvement efforts in our most high-risk environments are integrated into our company's overall strategy.

3. Over the past 3 years more of our business operations are in high-risk environments.

4. Over the past 3 years we have expanded our social engagement in high-risk environments.

5. Our firm frequently pursues socially responsible actions that come at the expense of profitability or competitiveness.

6. Our top leadership / C Suite supports our social impact initiatives in high-risk environments.

7. We monitor competitors to ensure that our CSR and social impact decisions do not lag behind peers.

8. We undertake social activities that are not captured in our reporting (e.g. sustainability or ESG reports, annual reports).

9. Our company has **accessible** channels (e.g., grievance mechanisms, whistleblower mechanisms, and feedback processes) through which people affected by our operations in high-risk environments can raise concerns.

10. Our company uses defined metrics or indicators to evaluate the social outcomes of our activities in high-risk environments.

11. (Withdrawn)

12. Regulatory requirements drive how much we engage on social issues in these spaces.

13. Regulatory requirements reduce our ability to be flexible and responsive on social ("S" in ESG) activities.

14. Political polarization around ESG and CSR in our headquarters country has made it harder to pursue social initiatives in high-risk environments.

15. Our company has clear criteria for deciding when to scale back or exit a region if conflict or instability escalates.

16. Information about security conditions from our local operations reaches senior decision-makers at headquarters in a timely and accurate way.

17. ESG ratings frameworks fairly capture our company's performance, activities, and risks in high-risk environments.

18. Local subsidiaries have autonomy to shape and implement social policies (as opposed to a centralized approach from HQ).

19. Our supply chain and sustainability teams engage regularly between each other.

20. My team / division engages regularly with senior management.

21. CSR at our firm is more relevant to our marketing than our operations departments.

22. Our operational presence can contribute to worsening social or conflict dynamics in some areas where we operate.

23. Demands from outside actors (investors, activists) make a tangible difference in our social impact strategy.

24. The European Union's Corporate Sustainability Due Diligence Directive (CSDDD) legislation has significantly changed our social activities for the better.